\documentclass[twocolumn,trackchanges]{aastex631}

\shorttitle{Spiral Disk Hosting Progenitor Globular Clusters in Perseus BCG}
\shortauthors{Lim et al.}
\graphicspath{{./}{figures/}}

\begin{document}

\title{Recent Formation of a Spiral Disk Hosting Progenitor Globular Clusters at the center of the \\ Perseus Brightest Cluster Galaxy: II. Progenitor Globular Clusters}

\author[0000-0003-4220-2404]{Jeremy Lim}
\affiliation{Department of Physics, The University of Hong Kong, Pokfulam Road, Hong Kong}
\email{jjlim@hku.hk}

\author[0000-0001-6491-9065]{Emily Wong}
\altaffiliation{Present address: Tokyo Institute of Technology, 2 Chome-12-1 Ookayama, Meguro City, Tokyo 152-8550, Japan}
\affiliation{Department of Physics, The University of Hong Kong, Pokfulam Road, Hong Kong}

\author[0000-0001-9490-3582]{Youichi Ohyama}
\affiliation{Academia Sinica, Institute of Astronomy and Astrophysics, 11F of Astronomy-Mathematics Building, No.1, Section\,4, Roosevelt Rd., Taipei 10617, Taiwan, R.O.C.}

\author[0000-0002-5697-0001]{Michael C. H. Yeung}
\altaffiliation{Present address: Max-Planck-Institut für extraterrestrische Physik, Giessenbachstraße, 85748 Garching, Germany}
\affiliation{Department of Physics, The University of Hong Kong, Pokfulam Road, Hong Kong}
\received{2021 September 8}
\revised{2022 January 11}
\accepted{2022 January 20}
\submitjournal{\apj}



\begin{abstract}
We address the nature and origin of Super Star Clusters (SSCs) discovered by \citet{star_clus_ID}  within a radius of $\sim$5\,kpc from the center of NGC\,1275, the giant elliptical galaxy at the center of the Perseus Cluster.  We show that, in contrast with the much more numerous population of SSCs subsequently discovered up to $\sim$30\,kpc from the center of this galaxy, the central SSC population have maximal masses an order of magnitude higher and a mass function with a shallower power-law slope.  Furthermore, whereas the outer SSC population have ages spanning a few Myr to at least $\sim$1\,Gyr, the central SSC population have ages strongly concentrated around $\sim$$500 \rm \, Myr$ with a $1\sigma$ dispersion of $\sim$100\,Myr.  These SSCs share a close spatial and temporal relationship with the ``central spiral," which also has a radius $\sim$5\,kpc centered on NGC\,1275 and a characteristic stellar age of $\sim$150\,Myr (Paper\,I).  We argue that both the central SSC population and the central spiral formed from gas deposited by a residual cooling flow, with the SSCs forming first followed by the formation of the stellar body of the central spiral $\sim$300--400\,Myr later.  The ages of the central SSC population imply that they are able to withstand very strong tidal fields near the center of NGC\,1275, making them genuine progenitor globular clusters.  Evidently, a spiral disk hosting progenitor globular clusters has recently formed at the center of a giant elliptical galaxy.

\end{abstract}



\section{Introduction} \label{sec:intro}

\begin{figure*}[htb!]
\centering
\includegraphics[width=\textwidth]{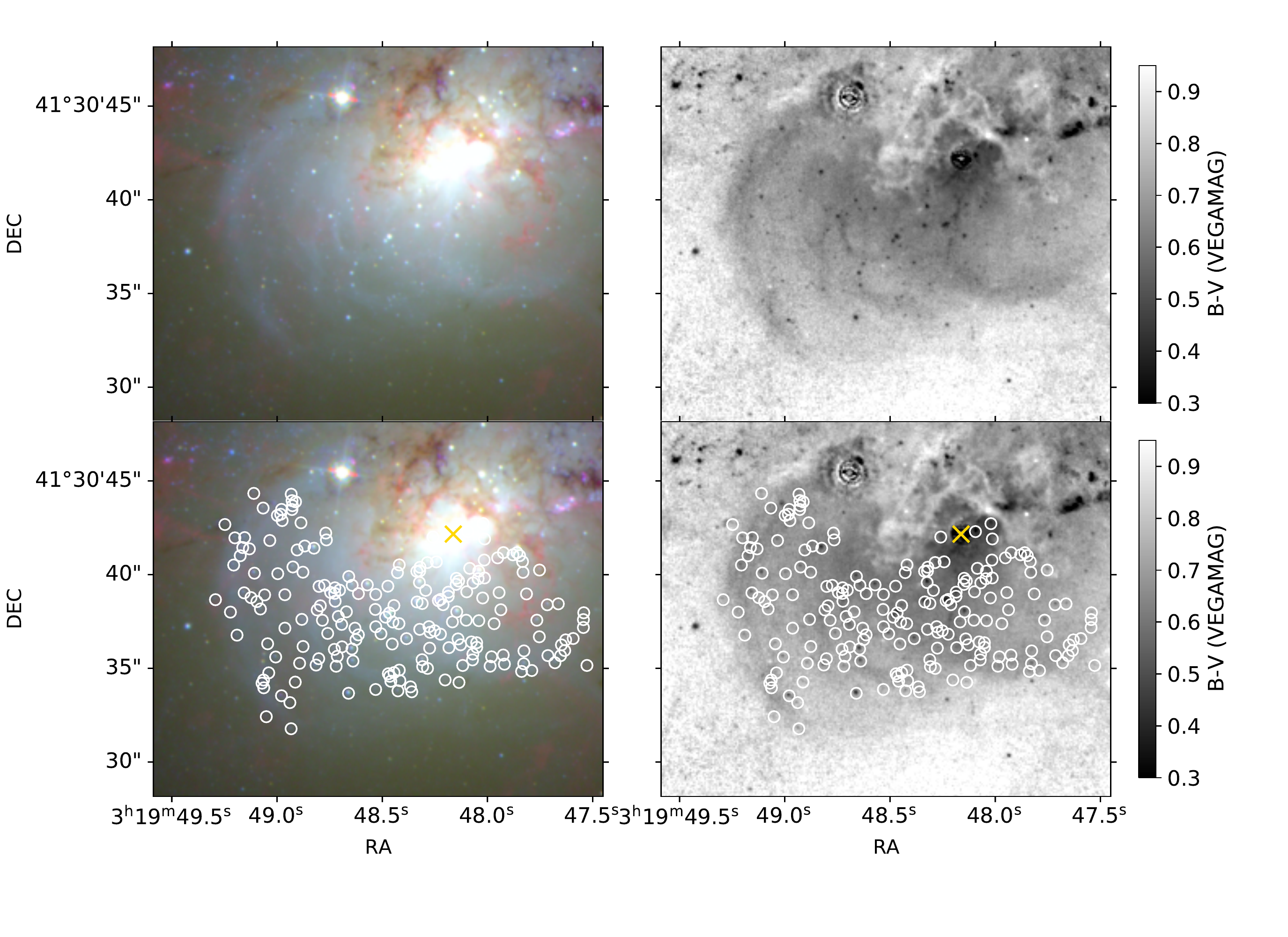}
\\
\vspace{-1.0cm}
\caption{$BVR$ (left column) and $B-V$ (right column) images of the central region of NGC\,1275 constructed from images in $B$ (F435W),$V$ (F550M), and $R$ (F625W) bands taken with the HST/ACS.  The $\times$ marks the position of the AGN in NGC\,1275.  The central spiral, comprising spiral arms superposed on an approximately circular disk, stands out in sharp relief owing to its higher surface brightness and bluer colors than its surroundings.  Filamentary dust seen in silhouette on the northern side of the galaxy belongs to the HVS, a distorted spiral galaxy falling toward NGC\,1275.  The SSCs selected for study are circled, comprising those projected primarily against the southern side of the central spiral free from overlap with the HVS.}
\label{fig:HST_image}
\end{figure*}

Apart from globular clusters (GCs), star clusters (SCs) in our Galaxy, as well as many other galaxies in the local Universe, have maximal masses of $\sim$$10^4 \, M_\sun$.  Some galaxies, however, possess super star clusters (SSCs) having masses of up to $\sim$$10^7 \, M_\sun$.  (Henceforth, we use the phrase SCs to refer to stellar clusters having masses of $\lesssim 10^4 \, M_\sun$, and SSCs to those having masses $> 10^4 \, M_\sun$.  We use the phrase stellar clusters interchangeably for either SCs, SSCs, or GCs.)  Despite their enormous masses, SSCs are very compact, with reported half-light radii no larger than $\sim$20\,pc and more typically $\lesssim 3$--5\,pc.  In both their masses and sizes, the vast majority of SSCs therefore resemble GCs, which also have masses of up to $\sim$$10^7 \, M_\sun$ and in most cases half-light radii of $\lesssim$10\,pc.  Interestingly, SCs, SSCs, and GCs appear to have similar power-law slopes for their mass functions (i.e., number of stellar clusters versus mass), suggesting a universal mechanism for the formation of stellar clusters over all mass scales \citep[see][]{Jeremy}.  (The mass functions of GCs reverse in slope below $\sim$$10^5 \, M_\sun$, attributed to disruptive processes that preferentially destroy lower-mass GCs.)

Although SSCs are oftentimes referred to as progenitor GCs, whether SSCs can survive to become GCs remains contentious.  A number of studies suggest an early short (up to only a few 10\,Myr) but highly disruptive phase that destroys a large fraction of SSCs \citep[e.g., see the review by][]{deGrijs2007}, most clearly and comprehensively demonstrated for the Antennae galaxies \citep[e.g.,][]{Fall2005,Mengel2005,Whitmore2007,Whitmore2010} comprising a pair of merging spiral galaxies.  Nonetheless, there are examples of SSC having ages up to at least $\sim$1\,Gyr
(e.g., \citealt{whitmore1993,Schweizer,Lim2013}; \citealt{Jeremy}) and even up to $\sim$5\,Gyr \citep[][]{Ostlin1998}, suggesting that (a fraction of) SSCs in some galaxies can long endure.  Mechanisms that can disrupt stellar clusters include stellar feedback (radiation and winds, both of which disperse the remaining gas that do not form stars), stellar mass loss, two-body relaxation, tidal shock, and tidal truncation, each of which can operate on a vastly different timescale.  The survivability of SSCs is expected to depend upon their individual star-formation efficiencies (i.e., mass of gas converted to stars), and once the remaining gas has been dispersed, then to their masses and sizes (at a given size, those having larger masses are more tightly bound and therefore less susceptible to dissolution), their orbits in a galaxy, and the tidal field of their host galaxies along their orbits.  These factors may well be different for individual SSCs in a given galaxy, as well as for the overall population of SSCs in different galaxies.

While the largest numbers of SSCs are generally found among interacting or merging starburst galaxies, SSCs also have been found in dwarf starburst galaxies as well as in apparently normal spiral galaxies.  SSCs have even been found in large numbers in an elliptical galaxy, NGC\,1275, the giant elliptical galaxy at the center of the Perseus cluster; i.e., the Perseus Brightest Cluster Galaxy (BCG).  Using the Hubble Space Telescope (HST), \citet{star_clus_ID} discovered about sixty SSCs within 5\,kpc of the nucleus of NGC\,1275, for which they inferred masses of $10^5$--$10^8 \, M_\sun$, unresolved diameters of $\lesssim$15\,pc, and ages of several hundred million years or less.  A modern HST image of these SSCs, on which the work reported in this manuscript is based, is shown in Figure\,\ref{fig:HST_image}.  Based on better-focused images taken by the HST, \citet{Carlson2001} measured an average half-light radius of 6.2\,pc (range from $\sim$1--100\,pc) for about 175 SSCs (spanning about 6 mag in $R$ band) over a region of size $\sim$$15 \times 15$\,kpc centered on the galaxy.  Since then, SSCs have been found in the thousands throughout NGC\,1275, located as far out as $\sim$30\,kpc (in projection) from the center of the galaxy \citep{Canning2010,Canning2014,Jeremy}.

\citet{Jeremy} have compiled a comprehensive list and made a detailed study of SSCs beyond 5\,kpc of the center of NGC\,1275.  They found that these SSCs have masses ranging from $\sim$$10^4 \,M_\sun$ (corresponding to the detection threshold) to $\sim$$10^6 \,M_\sun$.  Remarkably, the SSCs have formed at a nearly constant rate (at a temporal resolution of 100s\,Myr) for at least the past $\sim$1\,Gyr.  Many of these SSCs lie close to a complex, filamentary, and multiphase emission-line nebula in NGC\,1275; this nebula is attributed to a residual cooling flow,\footnote{Cooling of the intracluster X-ray emitting gas that is, in large part, re-heated by the action of an AGN in the BCG.  The manner by which a residual cooling of the intracluster gas nevertheless occurs is not fully understood, although likely to involve a complex interplay between AGN jets and the surrounding intracluster gas \citep[e.g.,][]{Qiu2020} rather than a simple inflow of cooling quiescent gas.} which therefore constitutes the gas reservoir for producing the SSCs.  

As mentioned in passing by \citet{Jeremy}, SSCs within $\sim$5\,kpc of the center of NGC\,1275 appear to form a separate population with different overall physical properties than those farther out: for one, the brightest of these SSCs are about an order of magnitude more luminous than the brightest SSCs farther out.  The physical properties of the central population of SSCs, and the relationship between this SSC population and the ``central spiral" \citep[the topic of Paper\,I;][]{paper1}, are the subjects of this manuscript.  In Section\,\ref{sec:physical properties}, we show that the SSCs projected against the central spiral have different physical properties and ages compared with those farther out.  In Section\,\ref{sec:central spiral}, we draw attention to the close spatial and temporal relationship between the central SSC population and stars in the central spiral.  In Section\,\ref{sec:origin}, we discuss the likely common or related origin of the central SSC population and the central spiral, before discussing whether these SSCs will likely survive the strong tidal fields at the inner regions of NGC\,1275.  Finally,  in Section\,\ref{sec:summary}, we provide a concise summary of our work and concluding thoughts. As in Paper\,I, we adopt $H_0 = 70 \rm \, km \, s^{-1} \, Mpc^{-1}$ so that at a redshift of $z = 0.01756$, the distance to NGC\,1275 is 74\,Mpc whereby $1\arcsec = 360 \rm \, pc$.

\section{Physical Properties of SSC\lowercase{s} in NGC\,1275} \label{sec:physical properties}

\begin{figure*}[bht!]
\centering
\gridline{\fig{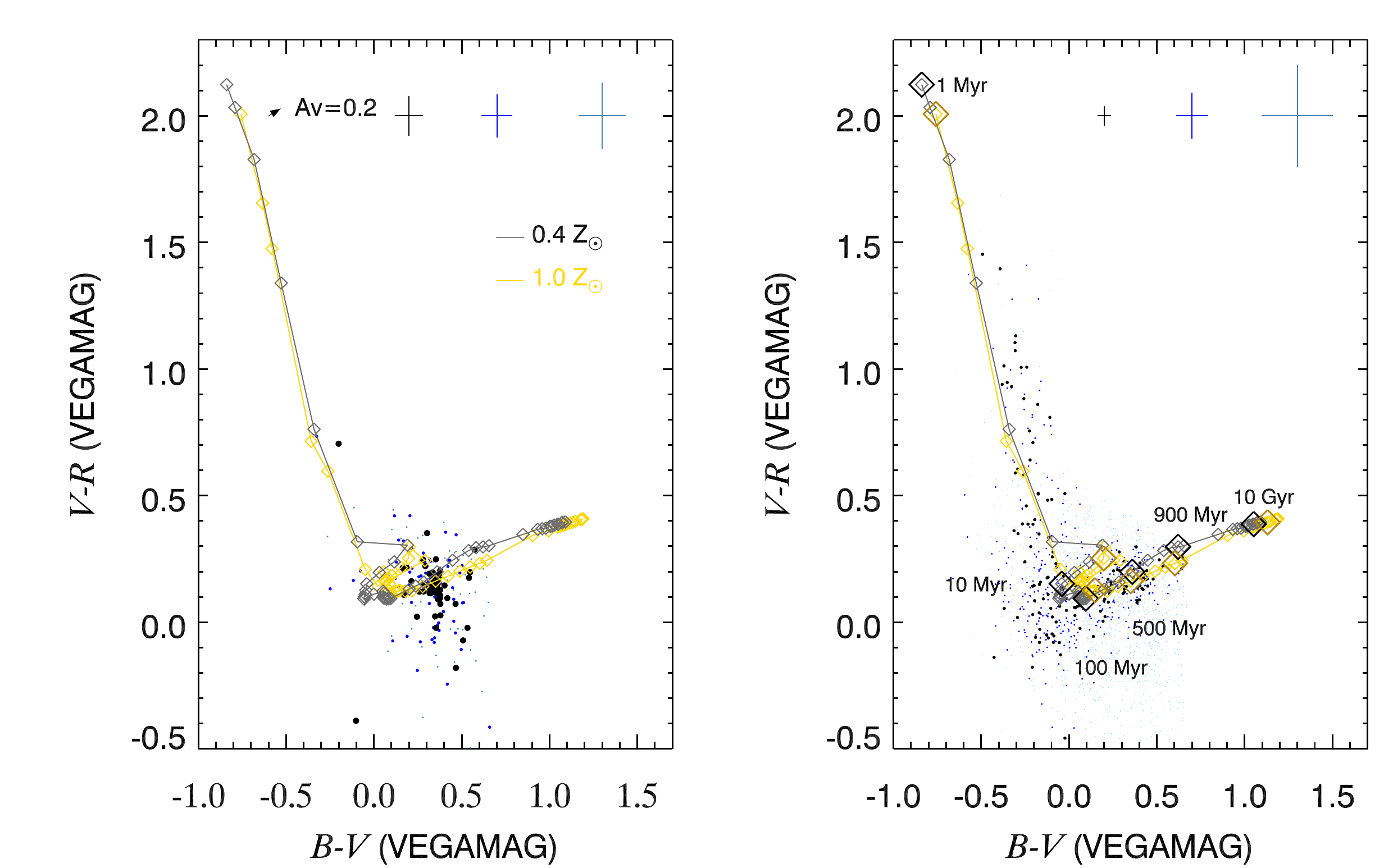}{1.0\textwidth}{}}
\vspace{-0.5cm}
\caption{$V-R$ versus $B-V$ of the SSCs cataloged by \citet{Jeremy}: (i) within a radius of $\sim$5\,kpc projected against the central spiral (left panel) as circled in Fig.\,\ref{fig:HST_image}; and (ii) beyond $\sim$5\,kpc from the center of NGC\,1275 (right panel) excluding those projected against the HVS and the outermost region dominated by GCs \citep[see Fig.\,2$a$ of][]{Jeremy}.  The SSCs are color-coded according to their typical measurement uncertainties in colors as indicated in the respective panels.  Notice that the SSCs projected against the central spiral are concentrated at $B - V \approx 0.35$ (see also Fig.\ref{fig:CM_diagram}), whereas those farther out span a broad range of colors (plot terminated where SSCs are overwhelmed in numbers by GCs).  Curves are age loci for a metallicity of $0.4 \, Z_\sun$ (black) and $Z_\sun$ (yellow).  Larger diamonds corresponding to ages as indicated, and smaller diamonds separated uniformly in logarithmic ages between the larger diamonds, are painted on the two age loci having different metallicities in the right panel.}
\label{fig:CC_diagrams}
\end{figure*}

Figure\,\ref{fig:HST_image} (left column) shows an image of the central region of NGC\,1275 constructed from observations with the HST/ACS in the broadband $B$ (F435W), $V$ (F550M), and $R$ (F625W) filters, and Figure\,\ref{fig:HST_image} (right column) shows the corresponding $(B - V)$ color image.  Spiral arms superposed on a roughly circular disk (see Fig.\,1 of Paper\,1 and also Fig.\,1 of \citealt{Carlson} for a larger field in which the entire disk is better seen), collectively referred to as the central spiral (Paper\,I), can be seen in clear relief owing to their relatively high surface brightness and blue colors by comparison with their surroundings.  The central spiral is closely centered on the nucleus of NGC\,1275, and extends outward to a visible radius of $\sim$5\,kpc.  A concentration of relatively luminous SSCs distributed throughout a similar radius of $\sim$5\,kpc as first reported by \citet{star_clus_ID} can be clearly seen.  Here, we select all 195 SSCs cataloged by \citet{Jeremy} in projection against primarily the southern half of the central spiral, corresponding to those circled in Figure\,\ref{fig:HST_image} (lower row), for detailed scrutiny\footnote{As described in \citet{Jeremy}, we selected candidate SSCs throughout the common field imaged by the HST in $B$, $V$, and $R$ by choosing objects that resemble the point spread function (PSF) of the telescope based on bright stars visible in the same images.  To this end, we used the StarFinder algorithm \citep{starfinder} that returns a correlation index ranging from 0 to 1, such that larger values correspond to a closer match with the PSF.  For candidate SSCs beyond a central radius of $\sim$5\,kpc, we imposed a correlation index of $> 0.67$.  For the more crowded and complex region with a central radius of $\sim$5\,kpc, we used a correlation index of $> 0.9$ so as to reduce contaminants.  Nonetheless, we verified that using the same correlation index of $> 0.67$ for both populations make no qualitative change to the results for the central population of SSCs as reported in this manuscript, in particular their different luminosity and mass functions by comparison to the SSCs lying farther out.}.  Those on the northern side are potentially contaminated heavily by stellar clusters belonging to the High Velocity System (HVS), constituting a distorted spiral galaxy moving at high speeds along the sightline toward NGC\,1275 \citep{HVS}.  A portion of the HVS is visible in Figure\,\ref{fig:HST_image} in the form of silhouette dust.

A color-color diagram in $V-R$ versus $B-V$ of the individual SSCs projected against the central spiral is shown in Figure\,\ref{fig:CC_diagrams} (left panel).  The manner in which photometry of the individual stellar clusters was performed, including background subtraction, is described at length in \citet{Jeremy}.  In brief, we used the StarFinder algorithm \citep{starfinder} to  perform PSF-fitting photometry that includes subtraction of the local background.  Only objects that exceed 4$\sigma_{\rm local}$, where $\sigma_{\rm local}$ is the local root-mean-square noise level, are accepted as detections.  Conversions to magnitude are based on the standard zero-points in the Vega system as described in the ACS Data Handbook \citep{ACS_Data_Handbook}, and corrected for Galactic extinction.  Two age loci, each of which is based on a model single stellar population (SSP; i.e., population of stars all having the same age and metallicity) as taken from \cite{ygg}, are plotted in Figure\,\ref{fig:CC_diagrams} for different metallicities of $Z = 0.4 \, Z_\sun$ (black) and $Z =  Z_\sun$ (yellow). The model SSPs are based on those utilized in the Starburst99 software \citep{starburst99_1,starburst99_2}, and their colors include both continuum and line emission from $\rm{H\,II}$ regions corresponding to an adopted unity filling factor for the gas left over from star formation.  The resulting line emissions, in particularly, H$\alpha+[\rm{N\,II}]$ that is contained in the $R$-band causes the steep rise in $V-R$ starting below $B-V \approx 0.0$ when the star clusters are younger than $\sim$10\,Myr. The two loci having different metallicities closely overlap, such that the inferred age at a given color differs little between the two loci.   The SSCs selected for study as indicated in Figure\,\ref{fig:HST_image} (lower row) are concentrated around $B-V \approx 0.35$, corresponding to an age around $\sim$500\,Myr (more details below).  The spread in their colors is larger than can be accounted for by photometric uncertainties alone, although there may well be a degree of contamination by stellar clusters associated with the HVS, as well as the chance projection of SSCs lying beyond 5\,kpc but positioned along the sightline toward the central spiral.  By contrast, as shown in the color-color diagram of Figure\,\ref{fig:CC_diagrams} (right panel), the SSCs located beyond the central spiral \citep[omitting the region occupied by the HVS and the outer regions of NGC\,1275 dominated by GCs, as described by][]{Jeremy} span a continuous range of ages from a few Myr to at least $\sim$1\,Gyr, beyond which they cannot be distinguished from the even more numerous GCs around NGC\,1275.

\begin{figure*}[htb!]
\centering
\gridline{\fig{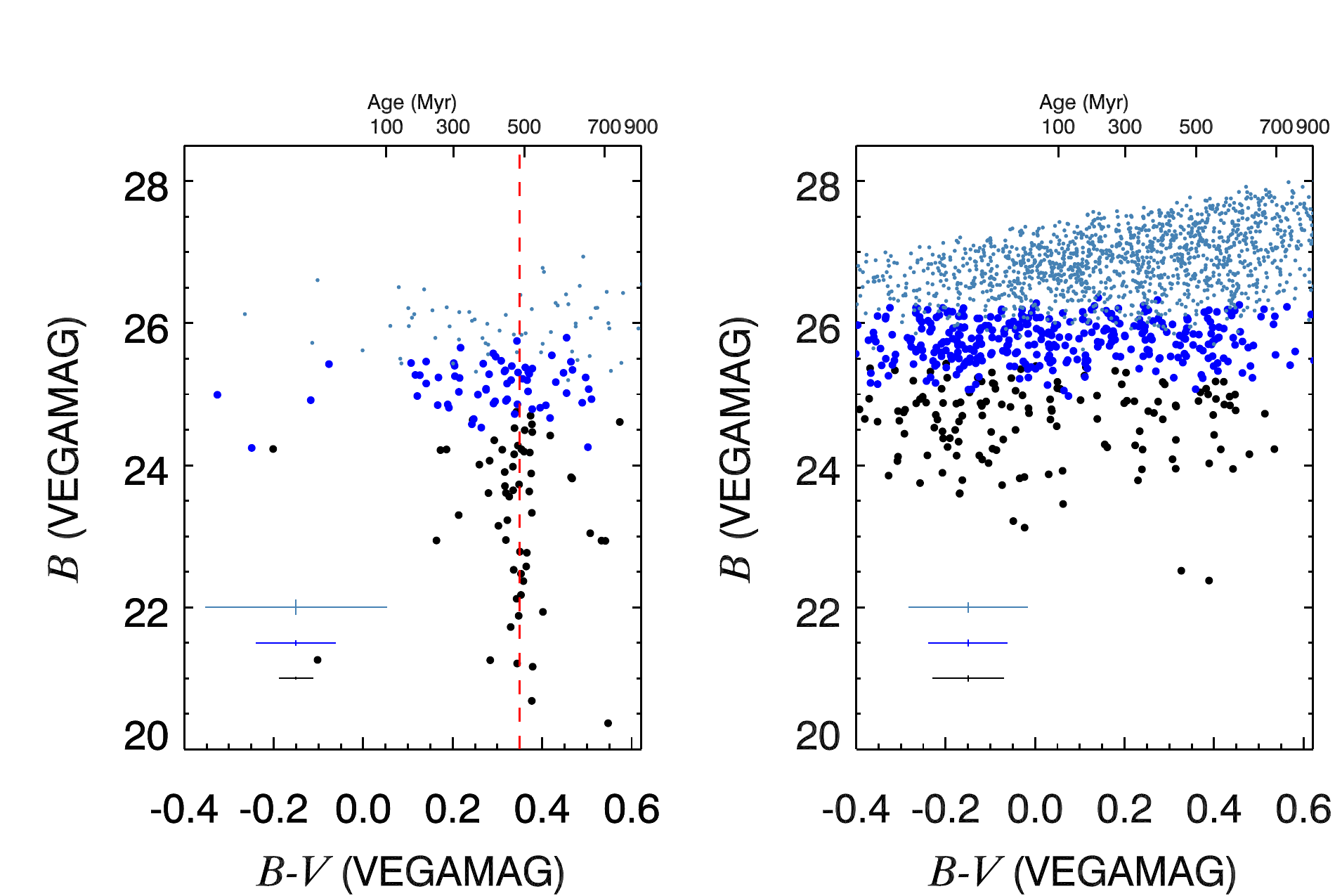}{0.9\textwidth}{}}
\vspace{-0.6cm}
\caption{Same population of SSCs as described in Fig.\,\ref{fig:CC_diagrams}, but now plotting their magnitudes in $B$ versus their colors in $B - V$.  Age as inferred from the $B - V$ color based on the age locus for $0.4 \, Z_\sun$ (see Fig.\,\ref{fig:CC_diagrams}) is indicated at the upper abscissa, applicable only for those with ages over 100\,Myr (see text).  The SSCs are color-coded according to their typical measurement uncertainties in colors as indicated in Fig.\,\ref{fig:CC_diagrams}.  Notice that the SSCs projected against the central spiral (left panel) are concentrated at $B - V \approx 0.35$ as indicated by the red dashed line, whereas those farther out (right panel) are quite uniformly distributed over a broad range of colors (this plot being terminated where SSCs are overwhelmed in numbers by GCs).}
\label{fig:CM_diagram}
\end{figure*}

Figure\,\ref{fig:CM_diagram} (left panel) shows a color--magnitude plot of the SSCs projected against the central spiral.  At an apparent magnitude in the $B$ band of $m_B \lesssim 24.5$, these SSCs are strongly concentrated at $B - V \approx 0.35$.  At $m_B > 24.5$, the SSCs have a broader color spread, albeit still closely centered at $B-V \approx 0.35$.  By contrast, as shown in the color--color diagram of Figure\,\ref{fig:CM_diagram} (right panel), the SSCs located beyond the central spiral (once again omitting the region occupied by the HVS and the outer regions of NGC\,1275 dominated by GCs) have a nearly uniform spread in colors irrespective of their brightness.  At the upper axes of both panels in Figure\,\ref{fig:CM_diagram}, we convert the measured $B - V$ color for each SSC to age based on the age locus of Fig.\,\ref{fig:CC_diagrams} for $Z = 0.4\,Z_\sun$.  Note that ages are not assigned to those having colors bluer than $B - V \approx 0.1$ (more strictly, those having ages below 100\,Myr), as these SSCs have similar $B - V$ colors, and ages assigned based on their $V - R$ colors are highly dependent on the luminosity of line emission produced by their $\rm{H\,II}$ regions that, in turn, is dependent on the unknown filling factor of gas left over from star formation.

\begin{figure*}[htb!]
\centering
\gridline{\fig{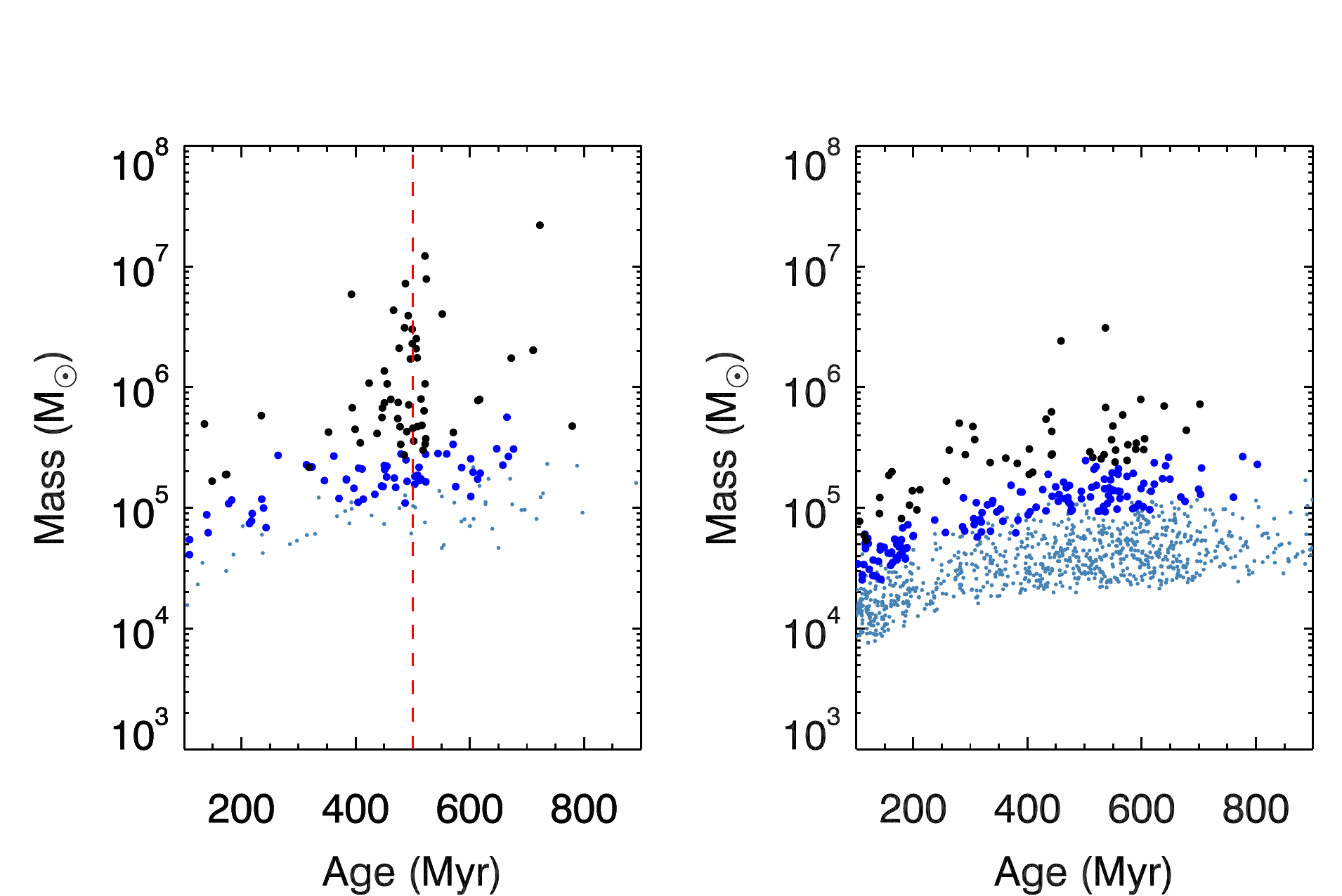}{0.9\textwidth}{}}
\vspace{-0.5cm}
\caption{Total initial stellar mass as inferred from the $B$ band luminosity versus age as inferred from the $B-V$ color for each of the SSCs plotted in Fig.\,\ref{fig:CM_diagram} having ages over 100\,Myr.  The SSCs are color-coded in the same manner as in Figs.\,\ref{fig:CC_diagrams}--\ref{fig:CM_diagram}.  Notice that the SSCs projected against the central spiral (left panel) are concentrated around an age of $\sim$500\,Myr as indicated by the red dashed line, whereas those located beyond the central spiral (right panel) are more uniformly distributed over a broad range of ages that is terminated at 900\,Myr in this figure.  Furthermore, those projected against the central spiral have maximal (total initial stellar) masses about an order of magnitude larger than those beyond the central spiral.}
\label{fig:age_mass}
\end{figure*}

\begin{figure*}[htb!]
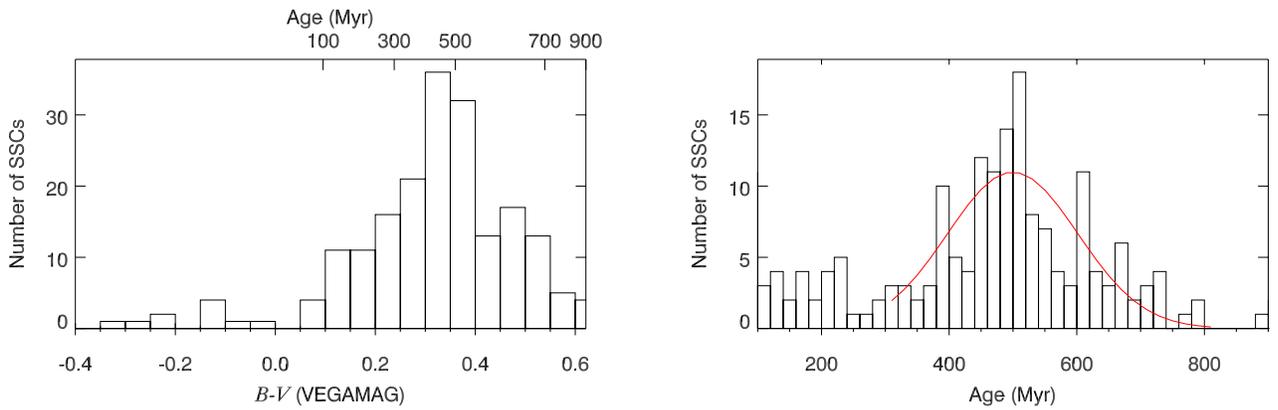

\centering
\gridline{\fig{histogram_B-V.pdf}{0.48\textwidth}{}
              \fig{histogram_color.pdf}{0.48\textwidth}{}}
\vspace{-0.8cm}
\caption{Number of SSCs projected against the central spiral versus $B - V$ color (left panel) and age (right panel) based on the age locus of Fig.\,\ref{fig:CC_diagrams} for a metallicity of $0.4 \, Z_\sun$.  The SSCs are concentrated at $B - V \approx 0.35$, corresponding to an age of $\sim$500\,Myr.  A Gaussian fitted to the histogram where the number of SSCs peaks is indicated by the red curve (spanning ages included in the fit), which is centered at $\sim$500\,Myr and a $1\sigma$ width of $\sim$100\,Myr.}
\label{fig:histogram}
\end{figure*}

In Figure\,\ref{fig:age_mass}, we convert the measured $B$-band luminosity for each of the SSCs having ages over 100\,Myr to its corresponding total initial stellar mass (i.e., at birth), and plot this mass versus age for each SSC.  The minimal (total initial stellar) masses of the SSCs projected against the central spiral are significantly higher than the minimal masses of those beyond, owing to the increased brightness of the background toward the center of NGC\,1275, and hence the higher detection threshold for star clusters within this region.   Whereas the SSCs beyond the central spiral span a broad range of ages terminated at 900\,Myr in this figure, the population projected against the central spiral are strongly concentrated around an age of $\sim$500\,Myr.  Furthermore, whereas the SSCs beyond the central spiral have maximal masses of $\sim$$10^6 \, M_\sun$, those projected against the central spiral have an order of magnitude larger maximal masses of $\sim$$10^7 \, M_\sun$.

\begin{figure*}[bht!]
\centering
\includegraphics[width=0.8\textwidth]{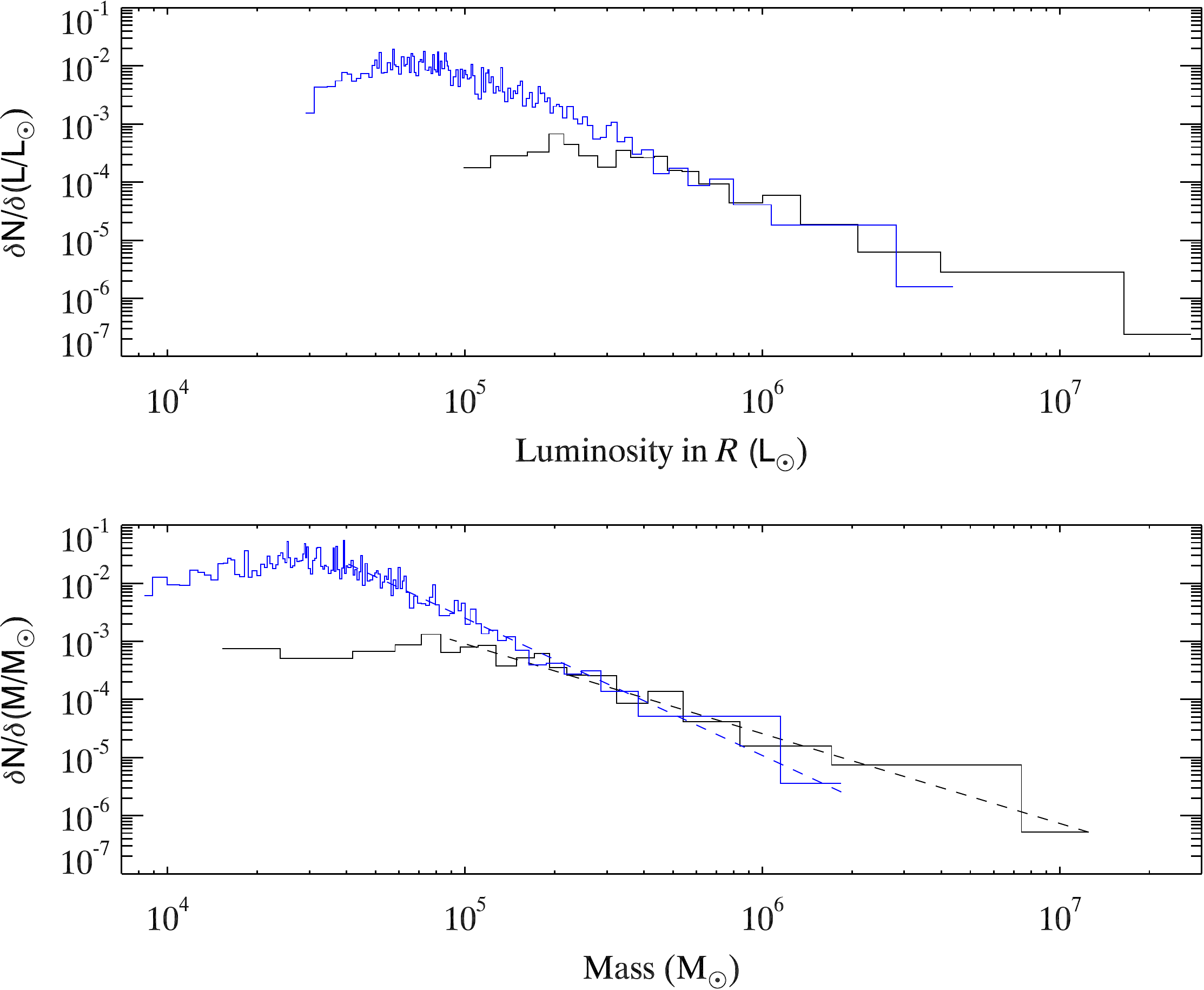}
\caption{Same population of SSCs as described in Fig.\,\ref{fig:CC_diagrams}, but now plotting their luminosity (upper panel) and mass (lower panel) functions in black for the central and blue for the outer population of SSCs.  Dashed lines indicate power laws fitted to the mass functions above their apparent turnover peak actually set by the observational detection threshold, and have different slopes of $-1.55 \pm 0.06$ for the central population of SSCs and $-2.37 \pm 0.06$ for the population of SSCs beyond.  Notice that SSCs projected against the central spiral have maximal luminosities and masses that are about an order of magnitude higher than those beyond the central spiral.}
\label{fig:luminosity_mass_functions}
\end{figure*}

Figure\,\ref{fig:histogram} (left panel) shows a $B - V$ color histogram of the SSCs projected against the central spiral.  As can be seen, the number of these SSCs peaks strongly at $B - V \approx 0.35$.  Figure\,\ref{fig:histogram} (right panel) shows an age histogram of the same SSCs, revealing that their number peaks strongly at 500\,Myr.  We fitted a Gaussian to this histogram around the age where the number of SSCs peaks, from which we infer an age centered at $\sim$500\,Myr and a $1\sigma$ width in age of $\sim$100\,Myr (note that this width arises from a combination of measurement uncertainties in color and therefore age, along with a true dispersion in age).  The inferred average age and age dispersion do not change significantly if we adopt the $Z = Z_\sun$ rather than the $Z = 0.4\,Z_\sun$ locus.  At ages below $\sim$300\,Myr, the SSCs exhibit a relatively uniform number distribution, but we cannot be certain whether these less luminous SSCs (see Fig.\,\ref{fig:CM_diagram}) constitute chance projection against the central spiral.

Figure\,\ref{fig:luminosity_mass_functions} shows the luminosity (upper panel) and mass (lower panel) functions of the SSCs projected against the central spiral (black), and for comparison, the luminosity and mass functions of the SSCs beyond the central spiral (blue) excluding the HVS and outermost regions dominated by GCs.  Despite possible contamination by unrelated SSCs that just happen to lie in projection against the central spiral, both the luminosity and mass functions of the SSCs projected against the central spiral have much shallower power-law slopes than the corresponding functions of the SSCs farther out.  In addition, the SSCs projected against the central spiral have maximal luminosities ($\sim$$10^7 \, L_\sun$) and masses ($\sim$$10^7 \,M_\sun$) about an order of magnitude higher than those farther out, despite being badly outnumbered and therefore much less likely to contain statistical outliers. As mentioned earlier, the higher detection thresholds for SSCs projected against the central spiral compared with those farther out is caused by the brighter background produced by the central spiral together with the old stellar population comprising the main body of NGC\,1275.

\section{Connection with the Central Spiral}\label{sec:central spiral}

As pointed out in Section\,\ref{sec:physical properties} and shown in Figure\,\ref{fig:HST_image}, the luminous population of SSCs within a central radius of $\sim$5\,kpc -- discovered by \citet{star_clus_ID} and studied in greater detail here -- is superposed on spiral arms imprinted on a roughly circular disk.  The nature of this central spiral has been studied in detail recently by \citet{paper1} (Paper\,I), who showed that: (i) the disk on which the spiral arms are imprinted is rotating quickly and decoupled from the main body of NGC\,1275; and (ii) its stellar population appears to be well-characterised by a single age of $\sim$150\,Myr (and likely bound within the range 100--200\,Myr) and an estimated total initial mass of $\sim$$3 \times 10^9  \, M_\sun$ (assuming solar metallicity).  This age, inferred from spectroscopic measurements that yield a higher accuracy and reliability than those inferred using only colors, is significantly younger than the average age of $500 \pm 100$\,Myr (note that the uncertainty is a combination of measurement uncertainties together with a true age dispersion) derived in Section\,\ref{sec:physical properties} for the SSCs projected against the central spiral.  Consistent with its younger stellar age, after separating the contribution from the dominant old stellar population comprising the main body of NGC\,1275 to the overall colors along the sightline toward the central spiral, we find the central spiral to have $B - V \approx 0.2$, which is bluer than the average color of the SSCs.

It will be of interest in Section\,\ref{subsec:origin} to compare the mass of stars in the central spiral with the total mass of SSCs projected against the central spiral.  The relatively shallow mass function of these SSCs implies that their total mass is dominated by the more massive stellar clusters.  Over the mass range $\sim$$10^5  \, M_\sun$ (the detection threshold) to $\sim$$10^7  \, M_\sun$ (the highest-mass SSCs), the total initial stellar mass of these SSCs is $\sim$$3 \times 10^8  \, M_\sun$, about one-tenth the initial stellar mass of the central spiral.  As we argue in Section\,\ref{subsec:survival}, those having lower masses are likely to have been preferentially disrupted, so that the mass function of these SSCs may have been much steeper in the past.  If their mass function at birth had been similar to that of the outer population of SSCs, then over the mass range $\sim$$10^5  \, M_\sun$ to $\sim$$10^7  \, M_\sun$ as before, the total initial stellar mass of the central SSC population would have been $\sim$$1 \times 10^9  \, M_\sun$, about one-third that of the central spiral.  

\section{Origin and Longevity of the SSCs}\label{sec:origin}
Despite the great strides made in our understanding of NGC\,1275 over time, the nature and origin of the extraordinarily luminous SSCs within the central $\sim$5\,kpc of NGC\,1275 have received little critical scrutiny since their discovery almost three decades ago by \citet{star_clus_ID}.  Previously in Section\,\ref{sec:physical properties}, we showed that these SSCs have different overall physical properties compared with those located farther out.  Then in Section\,\ref{sec:central spiral}, we showed that the SSCs located within the central $\sim$5\,kpc are not only spatially coincident with the central spiral, but are only somewhat older than the central spiral.  Their close physical and temporal relationships suggest a common or related origin.

\subsection{Origin}\label{subsec:origin}
\citet[][Paper\,I]{paper1} consider two possible origins for the central spiral.  First, it may be the remnant of a minor merger, constituting a (previously) gas-rich galaxy cannibalized by NGC\,1275.  If so, this merger remnant experienced a major starburst $\sim$150\,Myr ago that formed $\sim$$3 \times 10^9  \, M_\sun$ of stars.  This scenario could provide a natural explanation for the semicircular arcs around the central spiral visible out to a radius of $\sim$11\,kpc that \citet{conselice} and \citet{penny} attributed to the historical signature of a past merger.  The major difficulty with this scenario is the enormous mass in gas the merger remnant would have had to retain so as to fuel its recent starburst.  Even at an unrealistic nearly 100\% efficiency in converting gas to stars, the gas mass required is comparable to all of the molecular hydrogen gas in the Milky Way; for a star-formation efficiency of 1\%--10\% (the range that is more typically inferred for star-forming galaxies), the required gas mass is $\sim$$3 \times 10^{10}$--$3 \times 10^{11} \, M_\sun$.  It seems to us highly improbable that the progenitor galaxy could have retained such a large gas mass in its orbit close to the center of the Perseus cluster, during which time it would have experienced intense ram pressure stripping owing to the high density of the intracluster medium along with its high orbital velocity near pericenter.  This galaxy presumably made multiple orbits through pericenter before being slowed down sufficiently by dynamical friction---leading to a decay in both its pericentric and especially its apocentric distance---so as to merge with NGC\,1275.  This scenario would also require the merger remnant to have retained surrounding gas clouds so as to form the numerous SSCs associated with the central spiral.

Alternatively, in Paper\,I, we proposed that the gas for fuelling the recent starburst in the central spiral was deposited by a residual cooling flow.  Rather than being a merger remnant, perhaps a gaseous disk formed around the center of NGC\,1275 over time from a residual cooling flow, and then underwent a starburst $\sim$150\,Myr ago to form its present stellar body---seen as the central spiral.  This possibility is boosted by the presence of a more compact (radius of $\sim$100\,pc) rotating disk centered on the nucleus of NGC\,1275 \citep{scharwchter,Nagai} that is kinematically decoupled from the central spiral (see Paper\,I); this disk also was presumably deposited by a residual cooling flow.  In this scenario, the SSCs associated with the central spiral formed $\sim$300\,Myr before the starburst in the central spiral, presumably from gas clouds left over from the formation of the gaseous body of the central spiral.  Given the close synchronicity in time between the two events, the conditions that triggered the formation of SSCs around the central spiral may also have triggered the starburst in the central spiral -- indeed, the vast majority of SSCs discovered thus far have been born in close temporal association with, or themselves represent, starbursts.  

As mentioned above, both \citet{conselice} and \citet{penny} attributed the semicircular arcs around the central spiral to the historical signature of a past merger.  semicircular arcs comprising partial shells are quite commonly found around elliptical galaxies, where they are attributed to stars stripped from a lower-mass galaxy during a merger: the stripped stars spend longer times near the apocenters of their orbits where their radial velocities are close to zero, and therefore pile up at these locations to produce stellar overdensities.  The semicircular arcs around the central spiral, however, are relatively blue \citep{conselice}, suggesting that they are composed of relatively young stars rather than the disrupted body of a cluster member that might be expected to be composed primarily of relatively old stars (as the cluster member, largely stripped of gas, would have ceased star formation long ago).  Thus, rather than the historical signature of a past merger, the semicircular arcs around the central spiral may constitute stellar trails generated by the tidal disruption of SSCs that wandered too close to the center of NGC\,1275 along their orbits.

Finally, the formation of SSCs, rather than just SCs, is thought to be linked to the high external gas pressures in which their natal molecular clouds are immersed, an idea first proposed by \citet{Jog1992} and since widely invoked especially for the formation of SSCs in interacting and merging galaxies.  This idea could provide a natural explanation for why the SSCs associated with the central spiral have maximal masses that are an order of magnitude higher than those farther out: the inner regions of NGC\,1275 are presumably subjected to significantly higher gas pressures from the hot intergalactic and intracluster medium combined than the regions farther out.

\subsection{Survival Timescales} \label{subsec:survival}
The SSCs hosted by the central spiral are subjected to strong tidal fields near the center of NGC\,1275, and therefore reside in an extremely hostile environment for the dissolution of stellar clusters.  According to the theoretical simulations by \citet{Brockamp} for the survivability of newly formed globular clusters in present-day galaxies, strong tidal fields within a radius of $\sim$2.5\,kpc from the center of NGC\,1275 can potentially disrupt SSCs -- especially those having relatively low masses---over timescales of just several hundred Myr \citep[see discussion in Methods section of][]{Jeremy}.  Indeed, the SSCs hosted by the central spiral exhibit a shallower mass function than those farther out, suggesting that lower-mass SSCs within the central $\sim$5\,kpc have been preferentially disrupted.  Of course, the possibility of dissolution depends not just on the mass, but also the size, of the stellar cluster.  Consistent with this expectation, the SSCs within the central $\sim$5\,kpc of NGC\,1275 exhibit a range of smaller radii than those farther out \citep{Carlson2001}, suggesting that larger SSCs have been preferentially disrupted.  

In the theoretical simulations by \citet{Brockamp}, the early short-but-violent phase of strong tidal disruption  is followed by a temporally extended phase during which SSCs lose mass more gently owing to two-body relaxation.  During this phase, lower-mass stars are redistributed preferentially toward the outer regions of a stellar cluster and eventually escape.  The SSCs that remain around the central spiral therefore constitute those which can withstand the strong tidal fields near the center of NGC\,1275, and are losing mass only gently over time owing to two-body relaxation.  In just $\gtrsim 500$\,Myr, these SSCs will resemble GCs in their broadband colors, and they will be difficult to distinguish from genuine GCs.


\section{Summary and Conclusions}\label{sec:summary}
Our work provides a new understanding of the very luminous SSCs discovered by \citet{star_clus_ID} within a projected radius of $\sim$5\,kpc from the center of NGC\,1275, the central giant elliptical galaxy of the Perseus Cluster.  These SSCs have ages of $500 \pm 100 \,\rm Myr$ and maximal masses of $\sim$$10^7 \, M_\sun$, by comparison with the far more numerous population of SSCs lying farther out (at projected distances of up to $\sim$30\,kpc from the center) that have ages spanning the range from a few Myr to at least $\sim$1\,Gyr and lower maximal masses of $\sim$$10^6  \, M_\sun$.  Furthermore, both the luminosity and mass function of this central population of SSCs are shallower than those of the more spatially extended population of SSCs.

We draw attention to the close relationship in both spatial distribution and age between the central population of SSCs and the central spiral, which comprises spiral arms superposed on a roughly circular disk of radius $\sim$5\,kpc centered on the nucleus of NGC\,1275 -- the same spatial distribution as the luminous population of SSCs discovered by \citet{star_clus_ID}.  As shown by \citet[][Paper\,I]{paper1}, the disk on which the spiral arms are imprinted is rotating quickly and kinematically decoupled from the main body of NGC\,1275.  The central spiral is composed of or dominated by a stellar population having an age of $\sim$150\,Myr (and likely bound within the range 100--200\,Myr) and a total initial mass of $\sim$$3 \times 10^9 \,M_\sun$, implying that the central spiral experienced a starburst $\sim$150\,Myr ago.  Evidently, this starburst was preceded -- by a few hundred Myr -- by yet another starburst that produced the population of luminous and massive SSCs having a similar spatial extent as the central spiral and about one-tenth to perhaps one-third the total initial mass in stars.

Informed by the theoretical simulations of \citet{Brockamp} for the survivability of newly formed globular clusters in present-day galaxies, we argue that the SSCs hosted by the central spiral have survived an early short-but-violent phase of strong tidal disruption near the center of NGC\,1275, during which lower-mass and/or larger SSCs were preferentially disrupted.  This brief but highly destructive phase would naturally explain why the mass function of the central population of SSCs is shallower than the mass function of the SSCs lying farther out.  Furthermore, as shown by \citet{Carlson2001}, the SSCs within the central $\sim$5\,kpc of NGC\,1275 exhibit a range of smaller radii than the SSCs farther out, suggesting that larger SSCs close to the center of the galaxy also have been preferentially disrupted.  The remaining SSCs associated with the central spiral therefore constitute those that are able to withstand the strong tidal fields near the center of NGC\,1275, and are losing mass only gently over time owing to two-body relaxation.  In just $\gtrsim 500$\,Myr, these SSCs will exhibit broadband colors resembling GCs, and they will be difficult to distinguish from genuine GCs.

In conclusion, a spiral disk hosting progenitor globular clusters, visibly lacking only a bulge and halo, appears to have recently formed at the center of the BCG in the Perseus cluster.

\begin{acknowledgments}
This work was supported by the Research Grants Council of Hong Kong through a General Research Fund (17300620) to J.L., that also provided partial support for work by M.C.H.Y. toward an MPhil thesis.  Y.O. acknowledges the support by the Ministry of Science and Technology (MOST) of Taiwan through the grant MOST 109-2112-M-001-021-.  This research employed observations made with the NASA/ESA Hubble Space Telescope and made use of archival data from the Hubble Legacy Archive, which is a collaboration between the Space Telescope Science Institute (STScI/NASA), the Space Telescope European Coordinating Facility (ST-ECF/ESAC/ESA), and the Canadian Astronomy Data Centre (CADC/NRC/CSA).
\end{acknowledgments}

%

\vspace{5mm}
\facilities{HST (WFC ACS)}




\bibliography{sample631}{}

\bibliographystyle{aasjournal}


\listofchanges

\end{document}